\documentclass[twocolumn,showpacs,showkeys,preprintnumbers,amsmath,amssymb]{revtex4}

\usepackage{graphicx}

\usepackage{dcolumn}

\usepackage{bm}

\begin{document}




\title{\rm Microscopic description of  proton-induced spallation reactions with the 
           Constrained Molecular Dynamics (CoMD) Model }


\author{A. Assimakopoulou$^{1}$}

\author{G.A. Souliotis$^{1}$}

\email[Corresponding author. Email: ]{soulioti@chem.uoa.gr}

\author{ A. Bonasera$^{2,3}$ }
\author{ A. Botvina$^{4,5}$ } 

\author{ N.G. Nicolis$^{6}$ }

\author{ M. Veselsky$^{7}$ }

\affiliation{ $^{1}$
           Laboratory of Physical Chemistry, Department of Chemistry,
           National and Kapodistrian University of Athens, Athens 15771, Greece }
                     
\affiliation{ $^{2}$ Cyclotron Institute, Texas A \& M University, College Station, Texas, USA}    
          
\affiliation{ $^{3}$ Laboratori Nazionali del Sud, INFN, Catania, Italy }
     
\affiliation{ $^{4}$ Frankfurt Institute for Advanced Studies, Goethe University, 
                     D-60438 Frankfurt am Main, Germany }

\affiliation{ $^{5}$ Institute for Nuclear Research, Russian Academy of Sciences, RU-117312 Moscow, Russia}

\affiliation{ $^{6}$
           Department of Physics,
           The University of Ioannina, Ioannina 45110, Greece }           
           
\affiliation{ $^{7}$ Institute of Physics, Slovak Academy of
           Sciences, Bratislava 84511, Slovakia }


\date{\today}


\begin{abstract}


We studied the complete dynamics of the proton-induced 
spallation process  with the microscopic framework of the 
Constrained Molecular Dynamics (CoMD) Model.
We performed calculations of proton-induced spallation reactions on  
$^{181}$Ta, $^{208}$Pb, and $^{238}$U
targets with the CoMD model and compared the results with a standard two-step approach
based on an intranuclear cascade model (INC) followed by a statistical deexcitation model. The
calculations were also compared with recent experimental data from the literature.
Our calculations showed an overall satisfactory agreement with the experimental data and suggest
further improvements in the models. We point out that this CoMD study represents the first complete
dynamical description of spallation reactions with a microscopic N-body approach 
and may lead to advancements in the physics-based modelling of the spallation process.

\end{abstract}




\maketitle


\section{Introduction}

Spallation reactions induced by high-energy protons or neutrons are of vital importance
for both fundamental research and technical applications.
The most important areas of application of these reactions are 
the spallation neutron sources \cite{Clausen-2003,ESS}, the energy production techniques based on
accelerator driven systems (ADS) \cite{Daniel-1996,Nifenecker-1999}, 
the transmutation of radioactive waste \cite{Bowman-1998,Rubbia-1995,Nifenecker-2001,Wang-2016} 
and the radiation shield design for accelerators and space devices \cite{Koning-1998}.
Other areas of application include the development of radioactive ion-beams in ISOL-type 
facilities \cite{Lewitowicz-2011,RIB-2013} and the production of medical isotopes
\cite{Uyttenhove-2011,Abbas-2009}. 
The aforementioned applications require that  spallation observables be known  with high accuracy
in a broad range of projectile energies.
Many efforts have been devoted in obtaining experimental data on proton 
and neutron 
spallation reactions in the energy range (100--1000 MeV) with target materials 
used in the numerous applications. Because of the variety of the target nuclei 
and the wide range of energy,  empirical systematics approaches 
(see, e.g. \cite{SPACS,Prokofiev-2001,Ma-2017} and references therein) and theoretical modelling  
\cite{spallation_review,Xu-2016,Sharma-2017} with  satisfactory predictive power are indispensable.

Recently, systematic high-quality experimental data on spallation reactions were obtained
by the CHARMS collaboration at GSI, Darmstadt \cite{GSI}. In particular, the production of 
individual nuclides from charged-particle
induced spallation reactions were measured with high-resolution using the inverse kinematics
technique with the magnetic spectrometer FRS. 

In parallel to the various experimental efforts, theoretical developments have lead to 
advanced codes of the spallation process \cite{spallation_review}. 
However, there are still uncertainties concerning the  description of measured cross
sections and other observables. 
Traditionally, the spallation reaction is described as a two-stage process. 
The first (fast) stage involves a series of quasifree nucleon-nucleon collisions, the intranuclear
cascade, initiated by the incoming nucleon.
Several codes have been developed to describe this stage, such as the traditional INC code ISABEL
\cite{ISABEL}, the Li\`{e}ge Intranuclear Cascade Model INCL4.6 or INCL++ 
\cite{Boudard-2013,Mancusi-2014,Mancusi-2015}  and  the CRISP code \cite{Deppman-2013}. 
The second stage is described by a statistical deexcitation model, such as the binary decay code
GEMINI \cite{Mancusi-2010}, the evaporation-fission code ABRA07 \cite{Kelic-2008}
and the generalized evaporation model (GEM) \cite{Furihata-2000}. 
In the same vein, the Statistical Multifragmentation Model (SMM) 
\cite{Bondorf-1995,Botvina-2001,Botvina-2005} 
describes the compound nucleus processes  (evaporation-fission) at low energies 
and the evolution toward multifragmentation at high energies.

The primary goal of the present work is the description of the complete dynamics  of the 
spallation process using the microscopic Constrained Molecular Dynamics (CoMD) model  
\cite{Papa-2001,Papa-2005}.
We compared the CoMD results with recent experimental data from the literature
and with  two-stage calculations employing  the INC code ISABEL
followed by the deexcitation code SMM. 
In this work, we obtained mass yield curves, fission and residue cross sections and neutron multiplicities
for the  proton spallation of targets  $^{181}$Ta, $^{208}$Pb and $^{238}$U at 200, 500,and 1000 MeV.  
We chose these targets because of the availability of relevant  experimental data and the importance
of these materials in applications, especially  for accelerator-driven systems (ADS) and/or spallation neutron sources. 
The  paper is structured as follows: After a brief description of the theoretical models 
employed in this work, we present comparisons with experimental data from the literature.
A discussion and conclusions follow. Finally, in the Appendix, details of the CoMD 
procedure and the treatment of the surface term are described.


\section{Theoretical Models}

\subsection{Microscopic Model: CoMD}

The primary theoretical model employed in this work  enabling us to describe 
the complete process of spallation is the microscopic Constrained Molecular Dynamics
(CoMD) model originally designed  for reactions near and below the Fermi energy.
The present CoMD procedure is along the lines of our recent work on  
proton-induced fission \cite{Vonta-2015}.
We note that the CoMD  code is based on the gereral approach of molecular dynamics
as applied to nuclear systems \cite{Aichelin-1991,Bonasera-1994}. 
In the CoMD code \cite{Papa-2001,Papa-2005,Giuliani-2014}, nucleons are described as 
localized Gaussian wave  packets in coordinate and momentum space. The wave function  
of the nuclear system is assumed to be the product of these single-particle wave functions.
With this Gaussian description, the N-body time-dependent Schr$\ddot{o}$dinger equation
leads to (classical) Hamilton's equations of  motion for the centroids of the nucleon wavepackets. 
The potential part of the Hamiltonian consists of a simplified Skyrme-like   
effective interaction and a surface term. 
Proper choice of the surface term was necessary  to describe the fission/residue competition
\cite{Vonta-2015} as detailed in the Appendix. 

The isoscalar part of the effective  interaction corresponds 
to a symmetric nuclear matter compressibility of K=200 (soft equation of state). 
For the isovector part, several forms of the density dependence of
the nucleon-nucleon symmetry potential are implemented in the code.
Two of them were used in the present work, which  we called the standard potential 
and  the soft potential.
These forms correspond to a symmetry potential proportional to the density and 
its square root,  respectively  (see \cite{Papa-2013} and references therein).
We note that in the CoMD model, while not explicitly implementing
antisymmetrization of the N-body wavefunction, a constraint in the
phase space occupation for each nucleon is imposed,  restoring the Pauli principle
at each time step of the evolution. This constraint restores 
the fermionic nature of the nucleon motion in the nuclear system.
The short range (repulsive) nucleon-nucleon interactions are
described as individual nucleon-nucleon collisions governed by the
nucleon-nucleon scattering cross section, the available phase space
and the Pauli principle, as usually implemented in transport codes.
We point out that the present CoMD version fully preserves the total angular
momentum (along with linear momentum and energy) \cite{Papa-2005}, features
which are critical for the accurate description of the dynamics of fission
and/or particle emission.

In the present work, the CoMD code with its standard parameters
was used. 
The calculations were performed with both the standard  and the 
soft  symmetry potentials. 
The ground state configurations of the target nuclei
were obtained with a simulated annealing approach,  and were tested for stability
for relatively long times (1500--2000 fm/c).  These configurations were used in 
the CoMD code for the  subsequent simulations of the spallation reactions. 
For a given reaction, a total of approximately 10000 events were collected. 
For each event, the impact parameter of the collision was chosen in the range
b = 0--b$_{max}$ following a triangular distribution (to ensure, as usual, a uniform distribution 
on the surface of a circle with radius b$_{max}$). Guided by the experimental fission cross section
data and/or systematics, we chose the maximum impact parameter  b$_{max}$ to be 7.0, 7.5 and 8.0 fm 
for the three systems p+Ta, p+Pb and p+U, respectively. 
Each event was followed up to 15000 fm/c (5$\times$10$^{-20}$ s) and the phase space coordinates
were registered every 100 fm/c.  
At each time step, fragments were  recognized with the minimum spanning tree method 
\cite{Papa-2001,Papa-2005}  and their properties were reported. 
Thus, information on the evolution of the system and the properties of the resulting 
residues or fission fragments were obtained.
In this way, for fissioning events, the moment of scission of the deformed heavy nucleus
was determined. For these events, we allowed an additional time of t$_{decay}$=5000 fm/c after 
scission for the nascent fission fragments to deexcite and we analyzed their properties.


\subsection{Phenomenological two-stage description: INC/SMM}

Apart from the microscopic CoMD calculations, that constitute the primary goal 
of this work, we also performed 
calculations based on the customary two-stage scenario of the intranuclear cascade followed by
statistical deexcitation \cite{spallation_review}.

For the description of the intranuclear cascade stage, we used the code ISABEL \cite{ISABEL}
that we will simply refer to as INC in most of the instances in the following.  
This code is a well tested Monte-Carlo code with a long history of improvements. 
The target nucleus is simulated by a continuous medium bounded by a diffuse surface. 
Collisions between the incident nucleon and the nucleons of the target occur with a 
criterion based on the mean free path. 
Linear trajectories  are assumed between successive collisions.
Free nucleon-nucleon cross sections are used. 
The code allows for elastic and inelastic nucleon--nucleon collisions. Furthermore, it takes full 
account of Pauli blocking i.e. interactions resulting in nucleons falling below the Fermi
sea are forbidden. 
From a given INC event we obtain the mass number, the atomic number, the velocity, the 
excitation energy and the angular momentum  of the primary residue, as well as the kinematical parameters
of the accompanying nucleons.
The choice of the impact parameter follows the procedure previously described for CoMD.
For this work, 20000 INC events were generated for each reaction.

The deexcitation of the hot heavy fragments from the INC stage is performed with the 
Statistical Multifragmentation Model (SMM) 
\cite{Bondorf-1995,Botvina-2001,Botvina-2005,Souliotis-2007,Botvina-2013}.
The SMM code combines a description of sequential compound nucleus decay with a multifragmentation model
in which the effect of angular momentum is carefully considered.
More specifically, in SMM, all possible decay processes occurring in the wide excitation energy range 
realized in a spallation reaction  are taken into account.
We note that for the deexcitation of low energy  ($\epsilon^*$ $<$ 1.0 MeV/nucleon) 
non-fissionable nuclei \cite{Veselsky-2011,Fountas-2014}),
the SMM code has been shown to  adequately describe  the particle deexcitation process as 
a cascade of emissions of neutrons and light charged particles 
using the  Weisskopf-Ewing model of statistical evaporation.  
In regards to fission of heavier excited nuclei \cite{Botvina-2013,Vonta-2016},
the following approach is followed.
A "multifragmentation" threshold value of $\epsilon^*_{mult}$ = 2.0 MeV/nucleon
is defined,  above which the SMM statistical multipartition is applied.
This threshold value is, of course, lower than the true nuclear multifragmentation threshold
of $\sim$3 MeV/nucleon, but it is employed as a parameter to define when the 
SMM multipartition scheme will be  applied to the decay of the excited nuclear system.
Thus, for intermediate and high excitation energies ( $\epsilon^*$ $>$ $\epsilon^*_{mult}$ ),
fission is simply described as a special case of multifragmentation, i.e., 
the binary partition of the excited heavy nucleus.
However, at low excitation energy (  $\epsilon^*$ $<$ $\epsilon^*_{mult}$ ),
the fission channel is described in the spirit of the liquid-drop model
with deformation-dependent shell effects. The Bohr-Wheeler approach is used 
for the calculation of the partial fission width.
The method for obtaining fission mass distributions is described in detail in \cite{Botvina-2013}.
We briefly mention that 
along with a symmetric fission mode, two asymmetric fission modes are included 
with contributions  dependent on the fissioning nucleus and the excitation energy.
These contributions, 
are described by an empirical parameterization based
on analysis of a large body of experimental data.
In the present work, apart from the value of $\epsilon^*_{mult}$ = 2.0 MeV/nucleon
that we used in the calculations presented in the following, we also tested 
the values 1.5 and 3.0 MeV/nucleon, and we found that the value  of 2.0 MeV/nucleon 
provides a satisfactory description of the experimental data on
heavy residues, fission fragments and intermediate-mass fragments.


\section{Results and Comparisons}


The main objective of the present work is the description of the complete dynamics
of the spallation process with the microscopic Constrained Molecular Dynamics (CoMD) model, 
that we recently used to describe fission at low and intermediate energy \cite{Vonta-2015}.
We performed simulations of proton-induced spallation on targets of $^{181}$Ta, $^{208}$Pb
and $^{238}$U at energies of  200, 500 and 1000 MeV.
We compared the results of CoMD calculations with available experimental data  
and  calculations that we performed with the traditional two-stage phenomenology 
using the INC/SMM model framework.

\begin{figure}[h]   
\begin{center}
\includegraphics[width=0.35\textwidth,keepaspectratio=true]{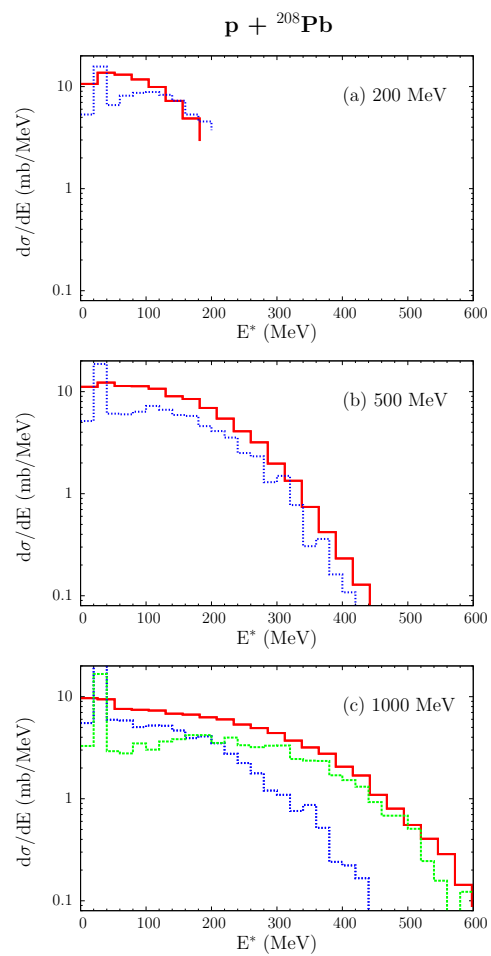}
\end{center}
\caption{ (Color online)
Excitation energy distributions of primary residues from proton-induced spallation of 
$^{208}$Pb at E$_{p}$=200, 500 and 1000 MeV calculated with the INC \cite{ISABEL} model
(solid red lines) and the CoMD model (dotted blue lines). In panel (c),  the dashed green line
is a CoMD calculation with nucleon-nucleon scattering cross sections increased by a factor of 2
(see text).
}
\label{figure01}
\end{figure}

In Fig. 1, we first present the excitation energy distribution of primary residues after the 
intranuclear cascade for the proton-induced spallation of the $^{208}$Pb target at 
E$_{p}$ = 200, 500 and 1000 MeV calculated with the INC model (full red lines in Figs. 1a, 1b, 1c,
respectively). 
From these distributions we obtain the mean excitation energies of the primary residues 
which are 70, 110 and 170 MeV   at the above proton energies, respectively.
These values are in overall agreement with the ones obtained by Cugnon et al. \cite{Cugnon-1997} 
with the INCL model (Fig. 3  of \cite{Cugnon-1997}). 
We remind that the INC code ISABEL has been extensively tested and shown to provide an overall
reasonable description of the properties of the primary residues. 
For representative comparisons, we first mention the recent work of Paradella et al. \cite{Paradella-2017}
on $^{136}$Xe (200 MeV/nucleon) + p. 
Moreover, Filges et al. \cite{Filges-2001} performed a comparison of primary residue excitation energy
distributions  calculated with ISABEL with experimental data from \cite{NESSY-expt} for  p(1.2GeV) + $^{197}$Au 
and saw a good agreement (Fig. 22 of \cite{Filges-2001}).
As the main focus of this work is on the microscopic description of the spallation dynamics,
we employed only the ISABEL code (referred to as simply INC in the following) as a representative code 
expected to provide  an overall satisfactory description of the intranuclear cascade stage of the reaction.

Moreover, we obtained excitation energies of primary residues with the CoMD code 
by simulating the intranuclear cascade stage with the time evolution of the system  
up to t=200 fm/c. (We also tested the times t=100 and 300 fm/c with comparable results.)
At this time, the binding energy of the excited residue was determined and compared 
with that of the corresponding ground state nucleus obtained from standard mass tables 
to get the excitation energy. 
The excitation energy distributions obtained from CoMD with the above approach
are shown in Fig. 1 by the dotted (blue) lines.
The CoMD distributions are in good agreement with the INC distributions for the
lower two energies, i.e. 200 and 500 MeV. However at 1000 MeV, the CoMD substantially
underestimates the residue excitation energy at the high end of the distribution.
We attribute this behavior to the fact that in CoMD no inelastic nucleon--nucleon collisions
are considered. In order to mimic the effect of inelastic collisions, we performed a CoMD 
calculation in  which the nucleon-nucleon (elastic) scattering cross sections were increased 
by a factor of 2. The resulting distribution (shown in Fig. 1c by the dashed green line) 
moved close to the INC distribution. 
We note that this deficiency of CoMD affects a rather small fraction of the events at the 
higher end of the excitation energy spectrum. In the present work,  we proceeded with
the standard values of the scattering cross sections as in the original code 
\cite{Papa-2001,Papa-2005}.

Finally, in the calculation of the excitation energy distributions with CoMD, we observe a concentration
of events at low excitation energy. We have seen an analogous behavior of the CoMD approach 
when applied to heavy-ion collisions \cite{Fountas-2014} leading to rather low (or even negative)
excitation energy for some of the primary products. In the future, we plan to improve the excitation
energy determination  by employing self-consistently calculated CoMD values of the ground-state
binding energies of  the primary residues,  so that we avoid the use of binding energies 
from mass tables.

As a general observation on the excitation energy distributions, the mean values 
are below the nuclear multifragmentation threshold of 2--3 MeV/nucleon. Only at the higher 
energy of E$_{p}$=1000 MeV, the tail of the distribution goes well above 2.0 MeV/nucleon
(400 MeV for a primary residue of typical mass A=200), thus corresponding to the onset
of multifragment emission \cite{Mancusi-2011}.
Similar observations pertain to the excitation energy distributions for the spallation
of the lighter Ta and the heavier U targets studied in this work.

In closing, we mention that in both the INC and the CoMD calculations on the cascade stage,
we have observed a close ralation of the mean residue mass with respect to excitation energy
that, due to the rather low excitation energy, does not depart substantially from the target
mass (it is lower, at most, by 5\%  for the higher excitation energy events at E$_{p}$=1000MeV). 
However, in spallation reactions with heavier projectiles and higher energies,
the mean residue mass gets progressively lower with increasing excitation energy,
as indicated in \cite{Botvina-1995} (Fig. 4 and 5) and the recent work \cite{Botvina-2017}
(Fig. 11).

\subsection{Mass yield distributions}
\begin{figure}[h]                                        
\begin{center}  
\includegraphics[width=0.45\textwidth,keepaspectratio=true]{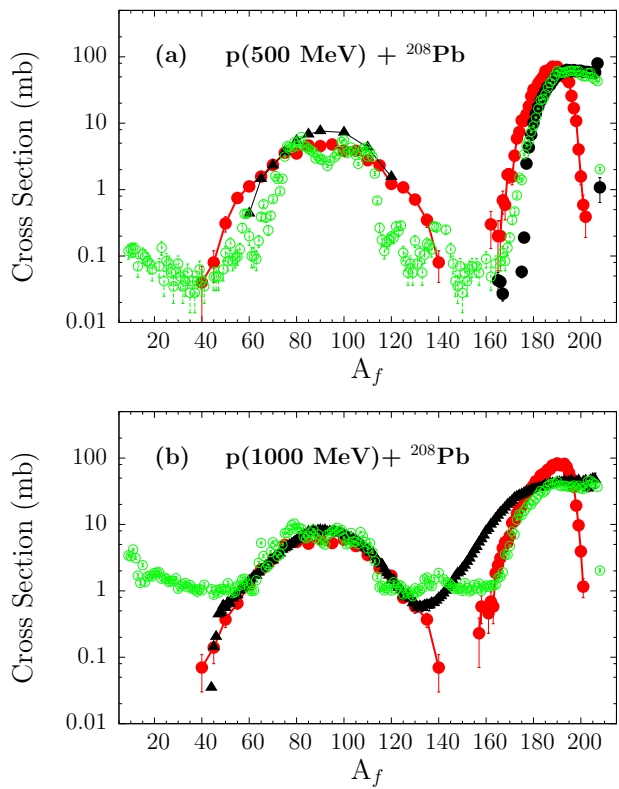}
\end{center}  
\caption{ (Color online)
Mass yield distributions  of fragments from  $^{208}$Pb spallation with protons at 500 MeV (a), and 1000 MeV (b).
Solid (red) circles: CoMD calculations. Open (green) circles: INC/SMM calculations.
Experimental data from the literature are as follows: in panel (a),  the solid (black) triangles are data on 
fission fragments from  \cite{Rodriguez-2015} and the solid (black) circles data on heavy residues from
\cite{Audouin-2006};
in panel (b), the  solid (black) triagles are data from \cite{Enqvist-2001}.
}
\label{figure02}
\end{figure}

In Fig. 2, we show the mass distributions of fragments from the  proton induced 
spallation of $^{208}$Pb at 500 and 1000 MeV.
We compare our theoretical results with  experimental data from the literature as follows.
In panel (a), we display the data of Rodriguez et al.  
\cite{Rodriguez-2015} (solid triangles) for fission fragments, and the data of Audouin et al.
\cite{Audouin-2006} (solid circles) for heavy residues. In panel (b), we show the data of
Enqvist et al. \cite{Enqvist-2001} (black triangles) for both fission fragments and heavy residues.
Our calculations with CoMD are presented by the solid (red) circles,  
and the INC/SMM calculations by the open (green) circles.
We note that the error bars on the calculation points in Fig. 2 (and all subsequent figures) 
are due to statistics.

As expected, in the mass distributions, we distinguish two regions of fragments. 
First, the  heavy residue region starting from  masses close to the target mass and extending
to lower masses, and, second,  the fission fragment region with a distribution of masses 
centered slightly lower than half  the mass of the target nucleus. 
As we expect, the shape of the fission fragment yield curve is symmetric, suggesting that
no significant contribution of low-energy fission processes is present in the reaction 
mechanism (that would lead to an asymmetric fission yield curve due to shell effects
\cite{Chaudhuri-2015}). 
We observe, that the CoMD calculated fission fragment yield curve is also symmetric
and in good agreement with the experimental data at both energies. 
We stress that the agreement is satisfactory not only in the shape, but also in the absolute
values of the fission cross sections.

The INC/SMM calculated fission yield curve shows a two-humped structure which is more
prominent in the lower energy reaction, Fig. 2a. The location and hight of these 
distributions are in  agreement with the data and the CoMD calculations, but the width is smaller.
A possible explanation for the double-humped structure of the fission yield curve from 
the INC/SMM calculation may be related in part to the parametrization of the low-energy
fission yield shapes adopted in the SMM code \cite{Botvina-2013}. 

In the heavy residue region the situation is more complicated. 
First, we see that the INC/SMM is able to reproduce the residue cross sections from the 
target mass down to approximately $\sim$180. For lower masses the INC/SMM calculation 
begins to largely underestimate the data at both energies.
This may be atributed to the contribution of very asymmetric binary decays
or multifragment decays that start to contribute appreciably 
as the excitation energy of the heavy remnant increases.

Furthermore, the CoMD calculation cannot reproduce the shape of the residue distribution. 
Our efforts so far have lead us to attribute this result mainly to the inability 
of CoMD to follow correctly the neutron-proton evaporation of a residue event, when we set
the surface parameter to the value appropriate to describe fission (see Appendix).
Moreover, we expect that in a residue event, the heavy fragment is still excited and would require 
substantially longer time to evolve (and, thus, give off the remainder of its excitation energy)
than the overall evolution time of 15000 fm/c in the present CoMD calculations.
Systematic efforts are being devoted at present to understand the above CoMD behavior 
and possibly achieve the simultaneous fission fragment and residue description. 
  
\begin{figure}[h]                                        
\begin{center}  
\includegraphics[width=0.45\textwidth,keepaspectratio=true]{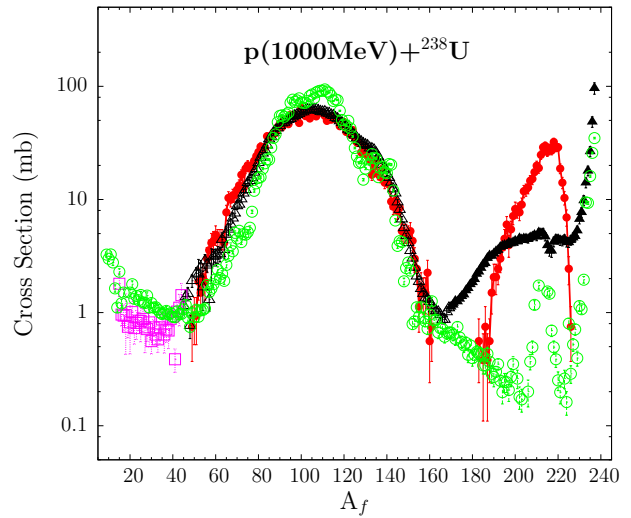}
\end{center}  
\caption{(Color online)
Mass yield distribution of fragments from $^{238}$U spallation with protons at 1000 MeV.
Solid  (red) circles: CoMD calculations. Open (green) circles: INC/SMM calculations. 
Experimental data from the literature are as follows:
solid (black) triangles: heavy residues from    \cite{Taieb-2003},
open (black) triangles:  fission fragments from \cite{Bernas-2003}, and 
open (pink) squares:     heavy IMFs from        \cite{Ricciardi-2006}.
}
\label{figure03}
\end{figure}

In Fig. 3, we present the mass distribution of proton-induced spallation 
of $^{238}$U  at proton energy of 1000 MeV.
Our theoretical results with CoMD [solid (red) circles] and INC/SMM [open (green) circles]
are compared with the following experimental data sets:
heavy residues from Taieb et al. \cite{Taieb-2003} [solid (black) triangles],
fission fragments from Bernas et al. \cite{Bernas-2003} [open (black) triangles] and 
intermediate mass fragments (IMF)  from Ricciardi et al. \cite{Ricciardi-2006} [open (pink) squares].
Observations similar to those in Fig. 2 pertain here.
As expected though, since the  $^{238}$U nucleus is more fissile, the fission yield curve is 
higher than in the $^{208}$Pb case (see also Fig. 5 and relevant discussion).
In the fission fragment region, interestingly, the CoMD calculations are in very good 
agreement with the experimental data.
The INC/SMM fission calculation gives a rather symmetric mass yield curve  in fair agreement 
with the CoMD calculation  and the data, having however, a smaller width. 

In the heavy residue region, the INC/SMM appears  to reproduce only the residue cross sections
close to the target mass. For lower masses the INC/SMM appears to mostly underestimate the data.
The CoMD calculation, as in the case of $^{208}$Pb (Fig. 2) cannot reproduce the 
shape of the residue distribution. 

Finally, apart from the heavy residue and fission fragment mass regions, 
we focus our attention on the region of heavy IMFs (A$<$50), for which the INC/SMM calculations
predict yields in reasonable agreement (slighly higher than) the data. 
Similar IMF products are obtained by the INC/SMM calculations for the p+Pb
spallation at 1000 MeV, Fig. 2c, for which no experimental data are available.
For the lower energy (500 MeV) p+Pb spallation (Fig. 2b), the INC/SMM calculations 
also predict IMF products with lower yields (by about a factor of 10).
We mention that we found the production of such fragments in CoMD in a small number
of events that involve  ternary (or possibly higher) partition of the heavy primary fragment. 
At 1000 MeV, the CoMD calculated cross sections are lower (by a factor of $\sim$5--10) 
than those obtained from the INC/SMM calculations. 
We understand that the observed IMFs are products of either very asymmetric fission or 
multifragmentation, and they may require further special efforts to be adequately described
(see also \cite{Colonna-2015,Pysz-2015} and references therein). 
This will be among the subjects of our future work.

In Figs. 2 and 3, we presented mass yield distributions for the reactions with Pb and U at energies 
where experimental data exist. For the p+Ta reaction, no mass yield distribution data are 
available at present (only fission cross section data exist and will be discussed below).
Moreover, we mention that our CoMD calculated yield distributions for p+Ta are similar in shape
to those of the p+Pb reaction at the corresponding energies.

From the study of the yield curves presented in Fig. 2 and 3, we see that the 
CoMD model appears to describe correctly the fission fragment yield curves
obtained in the proton-induced spallation of the $^{208}$Pb and $^{232}$U nuclei
at E$_p$ of 500 and 1000 MeV. However, in its present form it appears that it cannot describe
the residue mass distribution. 
For the INC/SMM calculation, we can say that some adjustment of relevant
parameters is necessary to improve the agreement with the data in the fission fragment region, 
as well as in the residue mass region.
Moreover, as demonstrated in Ref. \cite{Botvina-1990}, 
the excitation energies and nucleon composition of the excited
residues produced after the INC stage may require corrections
for preequilibrium emission.
As discussed above, systematic efforts are being devoted at present to understand 
the CoMD behavior for fission and residue events and possibly obtain a good description 
of both the fission fragment and the heavy residue distributions. 
Concerning the IMF products, appropriate adjustment of the elastic nucleon-nucleon 
scattering cross sections or, preferably, inclusion of inelastic channels 
is necessary for their description. 
 
\subsection{Fission and residue cross sections}

In Fig. 4, we present the variation of the fission cross section with 
respect to energy   for the proton induced spallation  
of $^{181}$Ta, $^{208}$Pb and $^{238}$U nuclei.
Our calculations were performed at energies 200, 500 and 1000 MeV. 
The CoMD calculations with the standard symmetry potential are depicted by
the full (red) circles connected with full lines, whereas those with the soft 
symmetry potential are shown by the full (blue) circles connected with dotted lines. 
The INC/SMM calculations are shown by the the solid (green) diamonds. 
In this figure, experimental data from the literature are presented as follows: 
In panel (a), we show the data of  Ayyad et al. \cite{Ayyad-2014} (solid triangles).
In panel (b), we show the data of Rodriguez et al. \cite{Rodriguez-2014} (solid triangles),  
Enqvist et al. \cite{Enqvist-2001} (open triangle), Schmidt et al. \cite{Schmidt-2013} (open square), 
Fernandez et al. \cite{Fernandez-2005} (open circle) and Flerov et al. \cite{Flerov-1972} (star).
In panel (c), we present the data of Kotov et al.  \cite{Kotov-2006} (solid triangles),
Bernas et al. \cite{Bernas-2003} (open square) and Schmidt et al. \cite{Schmidt-2013} (open diamonds).
Finally, in all panels the solid (black) line is according to the systematics of Prokofiev 
\cite{Prokofiev-2001}.

\begin{figure}[h]                                        
\begin{center}  
\includegraphics[width=0.40\textwidth,keepaspectratio=true]{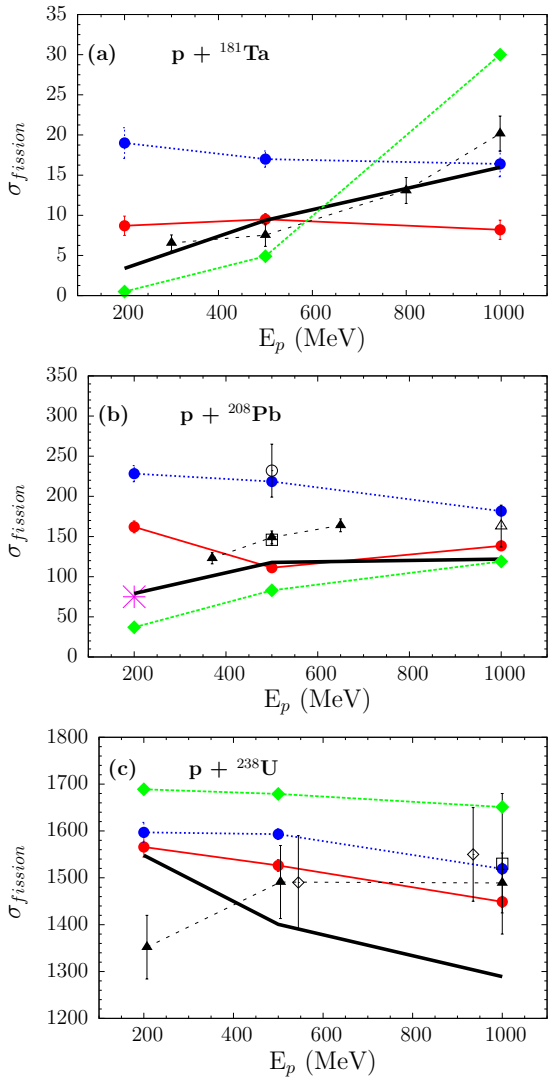}
\end{center}  
\caption{ (Color online)
Fission cross section as a function of proton energy for the proton-induced spallation 
of $^{181}$Ta,  $^{208}$Pb and $^{238}$U.  
Full (red)  circles connected with full lines: CoMD calculations with the standard symmetry potential. 
Full (blue) circles connected with dotted lines: CoMD calculations with the soft symmetry potential. 
Full (green) diamonds: INC/SMM calculations.
Experimental data from the literature are as follows. 
In panel (a), solid triangles from \cite{Ayyad-2014}. 
In panel (b), solid triangles from \cite{Rodriguez-2014}, open square from \cite{Schmidt-2013}, 
open circle from \cite{Fernandez-2005}, open triangle from \cite{Leray-2002} and star from  \cite{Flerov-1972}.
In panel (c), solid triangles from \cite{Kotov-2006}, open square from \cite{Bernas-2003} and
open diamonds from  \cite{Schmidt-2013}. In all panels, the thick solid line is according to the 
systematics of Prokofiev \cite{Prokofiev-2001}.
}
\label{figure04}
\end{figure}

The following observations pertain to this figure.
The experimental data and the empirical systematics show an increasing 
trend with energy for p+Ta and p+Pb, which is more pronounced for the less 
fissile p+Ta system (Fig. 4a).  This trend is roughly reproduced by the INC/SMM calculations, 
although with a steeper slope.
For the p+U, the data of Kotov et al. \cite{Kotov-2006} (solid triangles) show an increase 
from 200 to 500 MeV and then they stay constant for the higher energy. 
The higher energy flat behavior is also followed in the data of Schmidt et al. 
\cite{Schmidt-2013} (open diamonds).
Prokofiev's systematics, however, shows a decreasing trend. 
The INC/SMM calculations for this system show  a rather flat behavior and are 
higher than the data.

Furthermore, we observe that the CoMD calculations do not show an appreciable 
dependence on energy. (A rather slight decrease is discernible in Fig. 4b and Fig. 4c.)
The CoMD calculations with the soft symmetry potential are systematically higher than
those with the standard potential, implying that
the soft symmetry potential leads to a more repulsive dynamics in the neutron-rich 
neck region, as noted in our previous work \cite{Vonta-2015}.

For the p+Ta system (Fig. 4a),
the CoMD calculations with the standard symmetry potential are in agreement with the 
data at the lower energies, and those with the soft potential are in agreement with 
the data at the highest energy.
Focusing our attention on the p+Pb spallation (Fig. 4b), we observe that at the highest energy,
both the CoMD and INC/SMM calculations are in overall agreement with 
the data of Enqvist et al. \cite{Enqvist-2001} and the systematics of Prokofiev \cite{Prokofiev-2001}.  
At 500 MeV, the CoMD calculations are on either side of the data of 
Rodriguez et al. \cite{Rodriguez-2014} (solid triangle) and 
Schmidt et al. \cite{Schmidt-2013} (open square),
while the INC/SMM calculations are lower than the data and the CoMD calculations.
We note that the experimental point of Fernandez et al. \cite{Fernandez-2005} (open circle),
is higher that the data of Rodriguez et al. \cite{Rodriguez-2014} (solid triangle) 
and Schmidt et al. \cite{Schmidt-2013} (open square), and 
is close to the CoMD calculation with the soft symmetry potential.
Furthermore at 200 MeV, the CoMD calculations are higher than the INC calculations, whereas
the experimental point of Flerov et al. \cite{Flerov-1972} (star) and the systematics of 
Prokofiev \cite{Prokofiev-2001} (thick line) lie between these calculations.
Finally, for  the p+U spallation (Fig. 4c), we observe that 
at 1000 MeV, both CoMD calculations are in good agreement with the data of 
Bernas et al. \cite{Bernas-2003} (open square), 
Kotov et al.  \cite{Kotov-2006} (solid triangle) and 
Schmidt et al.  \cite{Schmidt-2013} (open diamond).
At 500 MeV, the CoMD calculations with the standard symmetry potential are in agreement with 
the data \cite{Kotov-2006,Schmidt-2013}, whereas at 200 MeV both CoMD calculations are higher 
than the data but in agreement with Prokofiev systematics (that starts to decrease with 
energy as noted previously).


\begin{figure}[h]                                        
\begin{center}  
\includegraphics[width=0.45\textwidth,keepaspectratio=true]{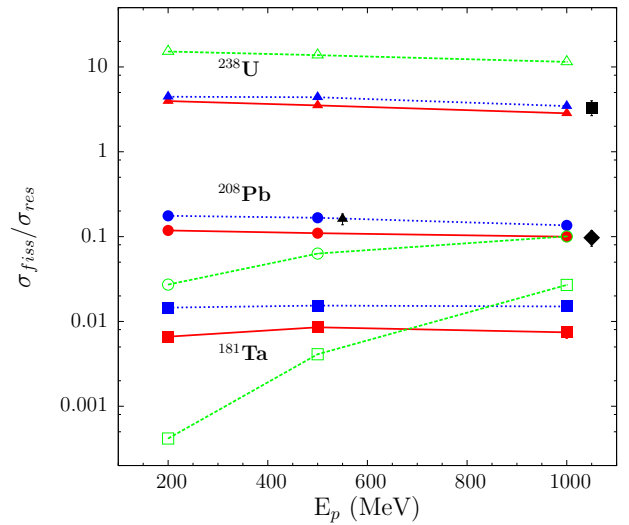}
\end{center}  
\caption{ (Color online)
Ratio of the fission cross section to residue cross section with respect 
to proton energy for the proton-induced spallation 
of  $^{181}$Ta, $^{208}$Pb, and $^{238}$U, nuclei.
The CoMD and INC/SMM calculations are with squares, circles and triangles for the above targets,
rspectively.
Full (red) points connected with full lines: CoMD (with standard symmetry potential). 
Full (blue) points connected with dotted lines: CoMD (with soft symmetry potential). 
Open (green) points connected with dashed lines: INC/SMM calculations.
The three experimental data points are: closed square \cite{Bernas-2003}, 
closed triangle \cite{Fernandez-2005}, 
and closed diamond \cite{Enqvist-2001}. 
The experimental points have been displaced by 50 MeV to the right for viewing purposes.
}
\label{figure05}
\end{figure}

In Fig. 5, we present the ratio of the fission cross section to residue 
cross section as a function of proton energy for the proton-induced spallation of 
$^{181}$Ta,  $^{208}$Pb, and $^{238}$U nuclei.  
The full (red) points connected with full lines are the CoMD calculations with the standard symmetry
potential. The full (blue) points connected with dotted lines are the CoMD calculations with 
the soft symmetry potential. 
The open (green) points connected with dashed lines are the INC/SMM calculations. 
In this figure, experimental data from the literature will be presented in the discussion
below.
(The experimental points have been displaced to the right by 50 MeV for viewing purposes.)

At first, we  observe that the CoMD calculation of the ratio for $^{238}$U is about 4, indicating, 
as expected, a high fissility. We notice also that the CoMD calculation at 1000 MeV is in good agreement
with the data of Bernas et al. \cite{Bernas-2003} (closed square). 
The ratio of fission cross section to residue cross section for $^{208}$Pb calculated with the CoMD
is about 10\% indicating the modest fissility of this target.  It appears that our
calculations are in good agreement with the data of  Fernandez et al. \cite{Fernandez-2005} at 500 MeV.
We note, however, that this experimental point involves a fission cross section value that is 50\% higher
than the trend of other experimental points and systematics, as discussed previously in regard to Fig. 4b.
At 1000 MeV, the CoMD calculations  are in good agreement with the data of 
Enqvist et al. \cite{Enqvist-2001}. 

Finally, for $^{181}$Ta, the ratio is only about 1\%, as calculated from CoMD.  This value suggests
that $^{181}$Ta has a very low fissility and thus, a tendency to undergo mostly evaporation.
There are no experimental data for the residue cross sections from the spallation of this target.
Only fission cross section data exist, that we presented in Fig. 4a. Thus, no experimental 
fission to residue cross section ratios are presented in Fig. 5.

The INC/SMM calculations [open (green) points] for p+U are higher than the CoMD calculations and the experimental 
point. For p+Pb they are lower than the CoMD calculations for the lower two energies and in agreement with the 
calculations and the data at the highest energy. A similar trend appears for the p+Ta system.

In general, in Fig. 5 we observe that the CoMD calculations with the soft potential are higher than those
with  the standard symmetry potential. We also oberved this trend in Fig. 4, in regard to the fission cross section.
Similar behavior was observed in our recent fission studies with CoMD \cite{Vonta-2015} and, as we mentioned before, 
was attributed  to the more repulsive dynamics implied in the neutron-rich neck region of the fissioning nucleus.


\begin{figure}[h]                                        
\begin{center}  
\includegraphics[width=0.40\textwidth,keepaspectratio=true]{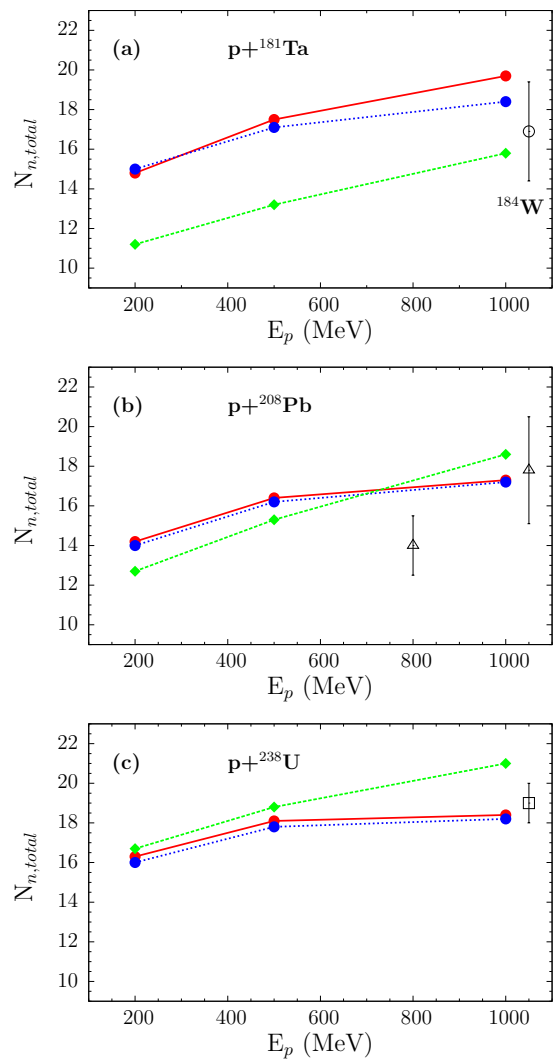}
\end{center}  
\caption{ (Color online)
Total Neutron Multiplicity as a function of proton energy for the proton-induced spallation 
of $^{181}$Ta,  $^{208}$Pb and $^{238}$U.  
Full (red)  circles connected with full lines: CoMD calculations with the standard symmetry potential. 
Full (blue) circles connected with dotted lines: CoMD calculations with the soft symmetry potential. 
Full (green) diamonds: INC/SMM calculations.
Experimental data from the literature are as follows:
In panel (a) and (b) from  \cite{Leray-2002}, and in panel (c) from \cite{Bernas-2003}.
}
\label{figure06}
\end{figure}

\subsection{Neutron multiplicities}

In Fig. 6, we present the total multiplicity of neutrons  emitted in the  proton induced spallation
of $^{181}$Ta, $^{208}$Pb and $^{238}$U at 200, 500 and 1000 MeV.
We note that the total neutron multiplicity refers to the number of neutrons emitted during 
the intranuclear cascade plus the number of neutrons evaporated during the deexcitation course 
of the hot primary residue. 
Specifically, if the deexcitation involves fission, the neutron multiplicity includes the 
prescission and postscission  neutrons.
Furthermore, we note that the total neutron multiplicity in  a fission event is nearly similar 
to that in a residue event. 
 
Our CoMD calculations of the neutron multiplicity refer only to the fission
events. We remind that, at  present,  the CoMD is not able to describe correctly the residue
evolution and, thus, the number of neutrons (and protons) given off by an evolving residue is 
incorrect. More specifically, the ratio of neutrons to protons given off by a residue  in 
CoMD (about 1--1.5) is much lower than the experimental value of approximately 4. 
In the figure,  the CoMD calculations with the standard symmetry potential are shown by 
the full (red) points connected with full lines, and those with the soft symmetry potential 
are shown by the full (blue) points connected with dotted lines. 
The two CoMD calculations are close to each other.

In the same figure, we present the results of the INC/SMM calculations [solid (green) diamonds].
In order to obtain the total multiplicity in these calculations, we added the neutrons emitted 
in the INC stage with the neutrons given off in the SMM deexcitation stage.
The INC/SMM results indicate an increasing trend with proton energy. They are 
lower than the CoMD calculations for p+Ta, moving closer for the p+Pb and p+U systems.

In Fig. 6a, the CoMD and the INC/SMM neutron multiplicity from the p+Ta spallation is in 
reasonable agreement with the experimental point of Leray et al. \cite{Leray-2002} (open circle)
for the p+$^{184}$W spallation at 1000 MeV. 
Similar agreement is seen in Fig. 6b between our calculations and the measurement
of Leray et al. \cite{Leray-2002} (open triangle) for the  p+Pb spallation. In this figure, 
however, the experimental point at 800 MeV is lower than the trend of our 
calculations.
Finally, in Fig. 6c, the CoMD calculation is in agreement with the value reported by 
Bernas et al. \cite{Bernas-2003} at 1000 MeV, while the INC/SMM calculation is somewhat 
higher.

As an overall observation, we notice that our calculations predict 
an increase of the spallation neutron multiplicity from 200 MeV to
500 MeV of proton energy. Above that energy, the multiplicity appears to increase  
rather slightly. Thus, our CoMD calculations verify that the proton energy region 500--1000 MeV
is a good choice for the effective generation of spallation neutrons.
In practice, as well known, spallation neutrons are produced in thick targets. 
Consequently, the whole energy region from the beam energy down to low energies contributes
to the spallation neutron spectrum. 
We thus appreciate the importance of understanding the evolution of the reaction mechanism
from low energy (50--100 MeV) up to the proton (or neutron) beam energy.
For spallation neutron yields from thick targets, we point to the relevant  references, 
e.g. \cite{Letourneau-2000}. 

Our successful calculation of spallation neutron multiplicity using the 
microscopic CoMD code  is very encouraging for our current efforts on the spallation
reactions. Given the  fully  dynamical  nature of the code, neutron energy spectra 
and angular distributions  can be obtained  and compared with relevant 
experimental data (e.g. \cite{Leray-2002}).
Similar comparisons can be performed for the multiplicities and the spectra of emitted protons 
and other  light charged particles in conjunction with recent experimental data
( e.g. \cite{Rodriguez-2016a,Rodriguez-2016b}).
This is the subject of ongoing systematic efforts in our group.
Furthermore we note that, along with the study of proton-induced spallation reactions, 
the study of neutron induced reactions is of special importance. However, systematic data
on these reactions are rather scarce (see e.g. \cite{Meo-2015} and references therein).
For such reactions, the predictive power of CoMD can be expoited, after appropriate 
benchmarking on the existing experimental data.

In closing, we wish to comment on the computational demand of our CoMD caclulations.
For a typical reaction, e.g. p (500 MeV) + $^{208}$Pb with the current implementation
of CoMD, we can generate approximately 30 events per day on a single core of a mid-range modern PC.
For a given reaction, we run on 10 independent processors for approximately one month to generate
10000 events that we subsequently analyzed.
The corresponding INC/SMM calculation is fast. For 10K events, the INC takes only a few minutes,
whereas the SMM deexcitation takes about 20 min.  
Thus the CoMD calculation is at least $\sim$10$^3$ times more computer-intensive than 
the typical fast INC/SMM  calculation. 
We may  conclude that, while the INC/SMM calculation can be used as an efficient event 
generator in transport codes for  practical applications, the microscopic CoMD approach can 
shed light on the dynamics of the spallation process and provide guidance to the tuning of the 
parameters of the phenomenological approaches employed for applications.

\section{Discussion and  Conclusions}  

In the present work we studied proton induced spallation 
of $^{181}$Ta , $^{208}$Pb, and $^{238}$U targets  in the energy range 200--1000 MeV.
We chose these  nuclei because of the availability of recent literature data and 
their significance in current applications of spallation.
We performed calculations of the full dynamics of the spallation process 
with the microscopic CoMD model. Moreover, we employed the traditional 
two-stage phenomenological approach based on the intranuclear cascade model (INC)
followed by deexcitation with the  SMM model.

Our CoMD calculations describe rather well the symmetric fission yield distribution 
in the spallation  of $^{208}$Pb and $^{238}$U targets, as concluded from the comparison 
with available experimental data. 
We encountered problems in the description of the residue yield distributions that 
we are currently investigating.
The INC/SMM calculations give a rather good description of the residues near the target,
but appear to give a narrrower  fission yield curve (whose shape is double-humped in the
case of the $^{208}$Pb spallation). We attribute this  behavior mainly to the 
empirical description of low-energy fission in the SMM model and we plan to investigate 
possible improvements. 
As suggested in \cite{Mancusi-2010,Botvina-1990}, among possible additions, the inclusion of a prequilibrium stage
between the INC stage and the deexcitation may lead to improvement of these calculations with 
the experimental data.

Our calculations reproduced rather well the total fission cross sections 
and the ratio of fission cross sections over residue cross sections, especially 
at the higher proton energy of 1000 MeV.
It is notable that the CoMD calculations do not reproduce the increasing trend of the 
fission cross sections with energy observed in the data. On the other hand, the INC/SMM 
calulations appear to give a steeper increase as compared to the data for the less fissile 
Ta and Pb targets and a nearly flat dependence for the U target.

Finally, our CoMD calulations appear to describe reasonably well the spallation neutron multiplicity
of spallation/fission events. The INC/SMM calculations  describe well the total neutron 
multiplicities for spallation events leading to fission or residue production at high energies.
In regard to the CoMD calculations, because of the fully  dynamical CoMD approach, energy spectra 
and angular distributions  of neutrons and light particles (e.g. protons, alphas)  can be obtained  
and compared with  experimental data.

It is desirable to extend our calculations with detailed comparisons using
experimental data on spallation reactions with the inverse kinematics
technique. Studies of residue and fission product isotopic and velocity
distributions of  low-fissility targets, such as $^{181}$Ta on hydrogen, 
could enhance the understanding of the spallation mechanism of non-fissile systems.
In this context, the role of facilities such as RISP \cite{RISP,RISP-2013} 
(currently in the design stages) will be vital.


Summarizing, we have shown that the microscopic CoMD code describes reasonably well  
the complicated many-body dynamics of the spallation/fission process. 
We wish to point out that the CoMD code provides results that are not dependent on the specific
dynamics being explored and, as such, it may offer valuable predictive power for the  spallation
observables. 
We plan to perform further systematic calculations of the observables of spallation reactions 
and compare them with available experimental data. These observables include, apart from the 
ones presented in this study, the energy distribution of the fission fragments, the isotopic
mass distributions of residues and fission fragments, as well as those of the heavy intermediate
mass fragments (IMFs). Finally, as we already mentioned, we plan to study the energy
and angular distributions of neutrons and light charged particles.
Along the lines of the present study, we expect that these efforts will shed light on the mechanism 
of the spallation process  and contribute to a quantitative physics-based description of spallation 
properties of  importance to applications.




\section{Acknowledgements}
We are thankful to  H. Zheng  and G. Giuliani for discussions on 
recent CoMD implementations.
We are also thankful to W. Loveland for enlightening comments
and suggestions on this work. 
Finally, we wish to acknowledge motivating discussions with 
S.C. Jeong, Y.K. Kwon, K. Tshoo and other members of the RISP facility. 
Financial support for this work was provided, in part, by
ELKE Research Account No 70/4/11395 of the National and Kapodistrian 
University of Athens. 
A. Botvina acknowledges the support of  HIC for FAIR (Germany).
M.V. was supported by the Slovak Scientific Grant Agency under contracts 2/0105/11
and 2/0121/14 and by the Slovak Research and Development Agency under contract
APVV-15-0225.



\appendix

\section{Adjustment of the surface parameter in CoMD}

In this Appendix, we give a description of the treatment of the surface term
of the effective interaction employed in the CoMD code in order to describe 
fission \cite{Vonta-2015}. First, we remind that the effective interaction usually 
employed in transport codes includes a surface term of the form:
\begin{eqnarray}
V^{\rm surf} &=& C_{surf} \nabla^{2} \rho
\end{eqnarray}
in which $\rho$ is the local density of the system. With the Gaussian wave-packet
description of the nucleons in CoMD, the form of the surface term becomes: 

\begin{eqnarray}
V^{\rm surf} &=& {C_{s} \over 2\rho_{0}} \sum_{i,j\neq i}
  \nabla^{2}_{\langle{\bf r}_{i}\rangle} (\rho_{ij})
\end{eqnarray}
in which $\rho_{ij}$ is the,  so called, superposition integral 
(or interaction density) \cite{Papa-2001,Vonta-2015}. 

The implementation of the surface term appears indispensible to correctly
reproduce the binding energy and the radius of a given nucleus.
In the CoMD code, the value of the surface parameter for heavy nuclei
(A$>$150) has been found to be $C_{s}$ = -2.0 \cite{Papa-2001,Papa-2005}
That is, this value is necessary to obtain a stable ground 
state configuration with the correct binding energy, radius and
collective properties (GDR, GQR, GMR  behavior) \cite{Souliotis-2010}.

In our previous work on fission \cite{Vonta-2015}, however, we found that we could not 
get fission  of U and Th nuclei with $C_{s}$ = -2.0. Instead, 
in order to correctly describe fission, the surface parameter had to 
change from its nominal value of $C_{s}$ = -2.0 to nearly $C_{s}$ = 0.0.
In the present work, a more careful adjustment was performed 
to reproduced the measured fission cross sections of the studied nuclei.

\begin{figure}[h]                                        
\begin{center}  
\includegraphics[width=0.40\textwidth,keepaspectratio=true]{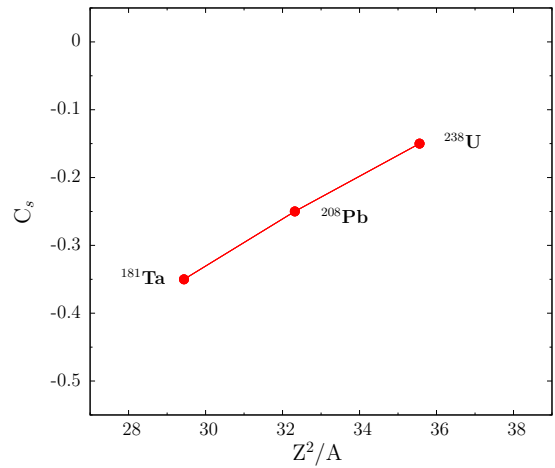}
\end{center}  
\caption{ (Color online)
The value of the surface parameter used in CoMD to describe the proton-induced 
spallation reactions of $^{181}$Ta, $^{208}$Pb and $^{238}$U  nuclei 
as a function of their fissility parameter Z$^2$/A  (see Appendix).
}
\label{figure_a01}
\end{figure}

The values of  $C_{s}$ that we found are plotted against the fissility
parameter of the respective nuclei in Fig. 7.
Interestingly, an overall linear trend of  $C_{s}$ w.r.t Z$^2$/A is 
observed. 
As already mentioned, a systematic study of the behavior of the surface term
(with respect to fissility and, possibly, excitation energy) is currently underway.


\newpage








\end{document}